\documentclass{ptephy}

\usepackage{bm,amssymb}
\graphicspath{./Figures/}

\begin{document}
\newcommand{\up}{\uparrow}
\newcommand{\down}{\downarrow}
\newcommand{\p}{\partial}

\title{Flow equation of functional renormalization group for three-body scattering problems}

\author{\name{\fname{Yuya} \surname{Tanizaki}}{1}}

\address{\affil{1}{Department of Physics, The University of Tokyo, Tokyo 113-0033, Japan}
\affil{1}{Theoretical Research Division, Nishina Center, RIKEN, Wako 351-0198, Japan}
\email{yuya.tanizaki@riken.jp}}

\begin{abstract}%
Functional renormalization group (FRG) is applied to the three-body scattering problem 
in the two-component fermionic system with an attractive contact interaction. 
We establish an exact flow equation on the basis of FRG and 
show that our flow equation is consistent with integral equations obtained from 
the Dyson-Schwinger equation. 
In particular, the relation of our flow equation and the Skornyakov and 
Ter-Martirosyan equation for the atom-dimer scattering is made clear. 
\end{abstract}

\subjectindex{A12,B32,D05}

\maketitle
\section{Introduction}
Few-body physics often provides fruitful information to study many-body physics. 
A typical example is the two-component fermionic system with an attractive contact interaction. 
At sufficiently low temperatures, the superfluid phase of this system shows crossover from the Bardeen-Cooper-Schrieffer (BCS) type superfluidity to the Bose-Einstein condensation (BEC) as the attraction becomes stronger \cite{PhysRev.186.456,leggett1980diatomic,nozieres1985bose,Zwerger201110}. 
In the BEC region, existence of composite bosons, or dimers, implies that low-energy excitations must be described by those dimers. 
Therefore, the scattering property between two dimers has to be taken into account for quantitative understanding of BEC (see e.g.  \cite{PhysRev.106.1135,PhysRevA.65.013606,PhysRevLett.83.1703}). 
Furthermore, these studies can be related to other systems, such as nuclear matter or dense QCD \cite{PhysRevC.73.044309,RevModPhys.80.1455}. 
In order to reflect knowledge of few-body physics in many-body theories, unified description for few-body and many-body physics is required. 
Since quantum field theory is a standard method for studies of many-body physics, it is important to establish a strategy of quantum field theory for few-body physics. 

Functional renormalization group (FRG) \cite{wetterich1993exact,Morris1,ellwanger1994flow}, which has been recently developed as a non-perturbative method of quantum field theory, provides a unified description between few-body and many-body physics. 
Indeed, it has been applied not only to study few-body physics in many different contexts 
such as three-body or atom-dimer scattering problems 
\cite{PhysRevC.77.047001,PhysRevC.78.034001}, 
trion formation \cite{PhysRevA.79.013603}, 
dimer-dimer scattering problems \cite{PhysRevA.81.043628,PhysRevA.83.023621}, and 
Efimov physics \cite{PhysRevA.79.042705,floerchinger2011efimov,schmidt2012efimov}, 
but also to study many-body physics, 
such as BCS-BEC crossover \cite{birse2005pairing,diehl2010functional,Boettcher:2012cm}.  
FRG is a formalism to calculate 
one-particle-irreducible (1PI) effective action $\Gamma[\psi]$ of fields 
$\psi,\overline{\psi}$
in a non-perturbative way. 
In this formalism, we define the 1PI effective action $\Gamma_k[\psi]$ by 
adding the two-point function $\overline{\psi}R_k\psi$ to the classical action $S[\psi]$, 
where $k$ is a parameter controlling the scale. 
Then the functional $\Gamma_k[\psi]$ obeys the exact one-loop flow equation 
\cite{wetterich1993exact,Morris1,ellwanger1994flow} 
\begin{equation}
\p_k\Gamma_{k}[\psi]
={1\over 2}\mathrm{STr}\left[\p_k{R}_k(\Gamma^{(2)}_{k}[\psi]+R_k)^{-1}\right], 
\label{wet01}
\end{equation}
where $\Gamma^{(2)}_k[\psi]$ is the second derivative with respect to the fields 
$\psi$ and $\overline{\psi}$. 
This equation is often called the Wetterich equation 
and holds an exact one-loop property. 
If the regulator term $R_k$ vanishes as $k\to 0$, then $\Gamma_k$ reduces 
to the usual 1PI effective action $\Gamma$, which contains all the information of 
quantum fluctuations. 

FRG must be consistent with the Dyson-Schwinger equation, since both methods 
provide exact relations among Green functions. 
Theoretically, we can prove that the Dyson-Schwinger equation can be regarded 
as an integrated equation of the Wetterich equation \cite{Pawlowski}. 
However, in previous studies 
\cite{PhysRevC.78.034001,PhysRevC.77.047001,PhysRevA.79.013603,
PhysRevA.81.043628,PhysRevA.83.023621,
PhysRevA.79.042705,floerchinger2011efimov,schmidt2012efimov} 
on few-body physics with FRG, 
flow equations in general contradict Dyson-Schwinger equations, 
which indicates that some contributions in the flow equations are missed. 
Flow equations in previous studies were constructed based on observations 
that the Wetterich equation has exact one-loop structure and 
particle-hole loop vanishes in the vacuum for non-relativistic physics. 
However,  some one-loop diagrams in the flow equation, 
which look like particle-hole loops at first sight, 
do not vanish even for non-relativistic scattering problems in the vacuum 
because of non-localities of 1PI effective vertices. 
As a result, a flow equation for $2n$-point 1PI vertex depends on $(2n+2)$-point 1PI vertex.     
This hierarchy of the flow equation can be solved in principle \cite{Floerchinger:2013tp}. 

In this study, we consider three-body problems for a contact and attractive interaction of two-component fermions, and establish the correct flow equation of FRG in a concrete way. In order to construct the closed flow equation without any approximations, we combine diagrammatic techniques with FRG so as to identify feedback from higher point vertices. Our flow equation is a closed equation and can be shown to be equivalent to the integral equation for three-body problem originally derived by Skornyakov and Ter-Martirosyan \cite{skornyakov1957three}. 
Through this example, we can observe relationship between flow equations and Dyson-Schwinger equations for few-body scattering problems. We also apply that relation to the model with an auxiliary dimer field in order to observe change of the structure of the flow equation due to introduction of the dimer field. 

Outline of this paper is as follows. In section \ref{two_body}, we fix our notation 
and briefly review the result on two-body scattering problem with an attractive 
contact interaction of two-component fermions. 
In section \ref{three_body}, we consider its three-body scattering 
problem using FRG without introducing auxiliary fields and analyze detailed structure of the flow equation. 
There we will find that feedback from higher point vertices do not necessarily vanish, 
however diagrammatic considerations will open a way to solve this hierarchy problem. 
As a result, we establish a new and correct flow equation for three-body problems in 
non-relativistic few-body physics. 
In section \ref{formal_sol}, we discuss the relation between our flow equation derived in 
section \ref{three_body} and Skornyakov and Ter-Martirosyan equation. 
This makes clear the connection between few-body physics in FRG and other 
conventional methods for scattering problems. 
In section \ref{auxiliary}, we briefly discuss the flow of few-body physics when an auxiliary dimer field 
is introduced, and suggest that our previous observation is quite general. 
Finally, we summarize our result and comment some perspectives in section \ref{conclusion}.

\section{Two-body scattering problem\label{two_body}}
Before going into three-body scattering problem, we need to consider and solve 
two-body scattering physics in the vacuum. 
We consider two-component fermionic systems with a contact interaction, 
where the bare classical Lagrangian of that model is given by 
\begin{equation}
\overline{\psi}\left(\p_{\tau}-{\nabla^2\over 2m}-\mu\right)\psi(x)
+g\overline{\psi}_\up\overline{\psi}_\down\psi_\down\psi_\up(x). \label{cl_ac}
\end{equation}
In (\ref{cl_ac}), $\psi$ represents a two-component fermionic field, $\overline{\psi}$ 
its conjugate, $\tau$ the imaginary time, $m$ the mass of particles, and $\mu$ 
the chemical potential. 
In order to treat scattering physics in this way, we need to put the system in the 
vacuum (the temperature $T=0$, and the particle number density $\rho=0$). 

In the effective action $\Gamma[\psi]$, all possible terms allowed by symmetries can 
appear in general. We introduce vertex functions as coefficients of the expansion 
in terms of the fields: 
\begin{eqnarray}
\Gamma[\psi]&=&\sum_{n=1}^{\infty}{1\over (n!)^2}
\int_{p_1,\ldots,p_n;q_1,\ldots,q_n}\hspace{-2.5em}
(2\pi)^{4}\delta^{4}(\sum_i p_-\sum_j q_j)
\nonumber
\\
&&\times
\Gamma^{\alpha_1\ldots\alpha_n}_{\beta_n\ldots\beta_1}(\{p_i\};\{q_j\})
\overline{\psi}_{p_1,\alpha_1}\cdots\overline{\psi}_{p_n,\alpha_n}
\psi_{q_n,\beta_n}\cdots\psi_{q_1,\beta_1}, \label{set08}
\end{eqnarray}
where $\int_p=\int{\mathrm{d}^4 p}/(2\pi)^4$, $\psi_p,\overline{\psi}_p$ are Fourier components of 
$\psi(x),\overline{\psi(x)}$, and $\alpha_i,\beta_i$ denote spin indices. 
Especially the four-point vertex $\Gamma^{\up\down}_{\down\up}$ consists of 
a spin-singlet part and a spin-triplet part due to spin-$SU(2)$ symmetry. 
Since the contact interaction does not have relative momentum dependence, only the spin-singlet 
part $\Gamma_k^S(P)$ survives, where $P$ denotes 
the center-of-mass momentum. 

\begin{figure}[t]
\begin{equation*}
-\p_k \Gamma_k^{S}(P)=\widetilde{\p}_k
\parbox{5em}{\includegraphics[width=3em]{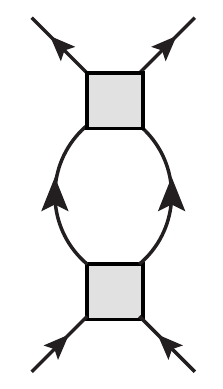}}
\end{equation*}
\caption{Diagrammatic expression of the flow equation of four-point couplings. 
Each line represents fermionic propagator and 
the square vertices represent $-\Gamma^S_k(P)$. }
\label{fig_fl4p}
\end{figure}

The flow equation for the four-point coupling $\Gamma_k^S(P)$ is 
\begin{equation}
-\p_k \Gamma_k^S(P)
=\widetilde{\p}_k \int_l{[-\Gamma_k^S(P)]^2\over [G_0^{-1}+R_k](l)[G_0^{-1}+R_k](P-l)}. 
\label{fl4p}
\end{equation}
Here we introduced the $k$-derivative $\widetilde{\p}_k$ which acts only on an explicit 
$k$-dependence of $R_k$ in propagators and $G_0^{-1}(l)=il^0+\bm{l}^2/2m-\mu$ 
is the free inverse propagator. Diagrammatic expression of (\ref{fl4p}) is given in 
Figure \ref{fig_fl4p}. 

Since the four-point couplings does not depend on relative momenta at the bare scale, 
relative momentum dependence of $\Gamma_k^S$ does not appear at any scale $k$. 
Renormalization condition of the four-point vertex is given by 
\begin{equation}
\Gamma^S(P^0=-2i\mu,\bm{P}=0)=4\pi a_S/m, 
\end{equation}
where $a_S$ is the $s$-wave scattering length. 
For an attractive contact interaction, the condition $a_S<0$ implies the absence of 
bound states and we can put $\mu=0$ in realizing $\rho=0$ at $T=0$. 
If $a_S>0$, there exists a bound state 
with the binding energy $1/ma_S^2$ and we should put $\mu=-1/2ma_S^2$ 
to realize $\rho=0$ and $T=0$. 

\begin{figure}[t]
\begin{equation*}
\parbox{3em}{\includegraphics[width=3em]{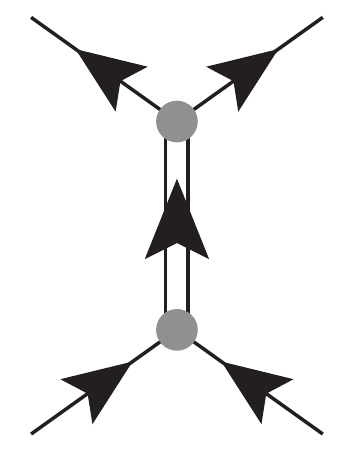}}
=\parbox{3.5em}{\includegraphics[width=3.5em]{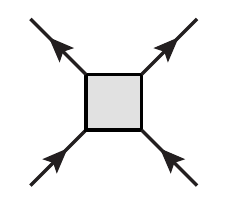}}
\end{equation*}
\caption{Identification of the four-point coupling $\Gamma^S(P)$ and the bound-state propagator $D(P)$. 
This diagrammatic identification is meaningful even without the bound state. }
\label{vac17}
\end{figure}

If the bound state exits, we can identify the boson propagator by looking at its pole in 
the effective vertex function. 
Then, we can regard the fermion four-point vertex as in Figure \ref{vac17}: 
In the left hand side the vertex represents $\sqrt{8\pi/m^2a_S}$ 
and the arrowed double line represents the rest of $\Gamma_k^S(P)$. 
If $a_S>0$, a bosonic bound state exists, and the double line at $k=0$ becomes 
the dimer propagator 
$D(P)=-\left(1+\sqrt{1+{m a_S^2}(iP^0+{\bm{P}^2\over 4m})}\right)/[2(iP^0+\bm{P}^2/4m)]$. 

The flow equation of $\Gamma_k^S$, (\ref{fl4p}), and graphical notations introduced 
in Figures \ref{fig_fl4p} and \ref{vac17} will be used in following sections. 

\section{Three-body scattering problems in FRG\label{three_body}}
Let us go into three-body scattering problems. In case the bosonic bound state exists, 
calculating six-point 1PI vertices is necessary for solving atom-dimer scattering problem 
in purely fermionic framework. 
The following discussion works well even when dimer does not exist. 

\begin{figure}[t]
\begin{equation*}
\p_k \parbox{3em}{\includegraphics[width=3em]{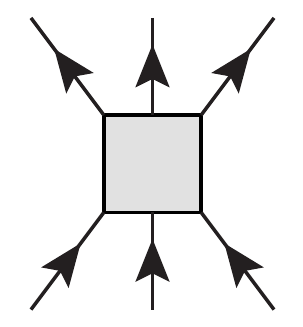}}
=\widetilde{\p}_k \left(\parbox{18em}{\includegraphics[width=18em]{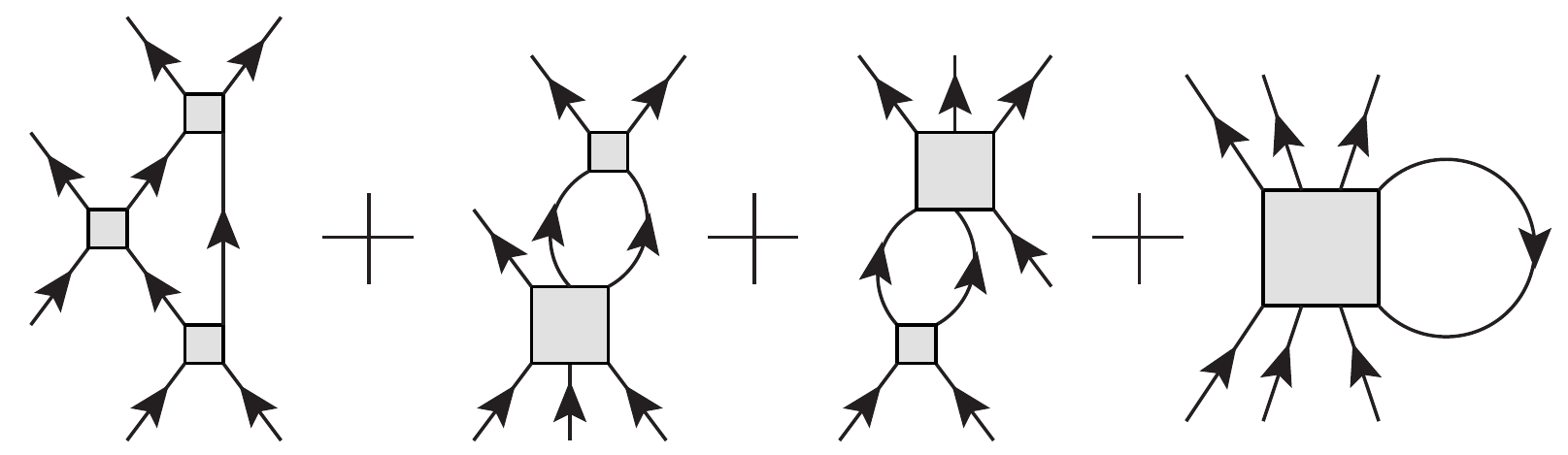}}\right)
\end{equation*}
\caption{Flow equation of the six-point 1PI vertex. Each square vertex represents 
negative of corresponding 1PI vertex. 
External legs of each diagram are totally antisymmetrized. }
\label{vac20}
\end{figure}

Let us start with the flow equation of a six-point 1PI vertex 
${\Gamma_k}^{\up\up\down}_{\down\up\up}$. 
The diagrammatic expression of its flow equation can be found in Figure \ref{vac20}, 
and its analytic expression will be given in Appendix \ref{app:6p}. 
There are two other diagrams in the right hand side of Figure \ref{vac20} 
when we expand the Wetterich equation (\ref{wet01}) in terms of fields 
$\psi,\overline{\psi}$, 
but diagrammatic consideration shows that they vanish because they necessarily contain 
particle-hole loops. 
The last diagram in the right hand side of Figure \ref{vac20} may seem to vanish at first sight 
since it also looks like a particle-hole loop, 
but it can contribute due to non-local dependence of effective vertices on energies. 
We will find in the next section that the flow equation for the three-body problem 
can be consistent with the Dyson-Schwinger equation 
only when this term is taken into account. 

If we use the flow equation in Figure \ref{vac20}, 
it depends on eight-point 1PI vertices 
and the flow equation of eight-point vertices again depends on ten-point vertices. 
In this way, the hierarchy of the Wetterich equation (\ref{wet01}) appears 
even for non-relativistic few-body problems. 
On the other hand, as a general property of non-relativistic few-body physics, 
we can solve $n$-body problems without considering $(n+1)$-body problems. 
These two seemingly-conflicting statements can be consistent, and the diagram 
containing eight-point 1PI vertices in Figure \ref{vac20} turns out to be written down 
within three-body scattering matrix elements as we will see. 

\begin{figure}[t]
\begin{equation*}
\parbox{3em}{\includegraphics[width=3em]{fig3_1}}=
\parbox{3.5em}{\includegraphics[width=3.5em]{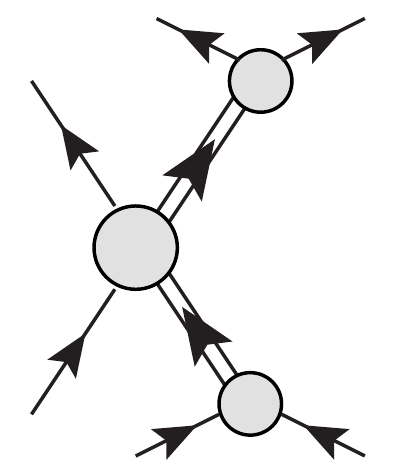}}
\end{equation*}
\caption{Decomposition of the six-point vertex ${\Gamma_k}^{\up\up\down}_{\down\up\up}$ by 
extracting external two-body scattering part $\Gamma_k^S$. }
\label{vac22}
\end{figure}

It is convenient to introduce 
the decomposition of the six-point 1PI vertex 
${\Gamma_k}^{\up\up\down}_{\down\up\up}$ by amputating external two-body $T$-matrices. 
Diagrammatic representation of this decomposition is given in Figure \ref{vac22}, 
and according to this 
decomposition we define a new vertex function $\lambda_k$ so that 
\begin{eqnarray}
&&{\Gamma_k}^{\up\up\down}_{\down\up\up}(p_1,p_2,p_3;p'_3,p'_2,p'_1) \nonumber\\
&&=
-\Gamma_k^S(p_{2+3})\lambda_k(p_1,p_{2+3};p'_{2+3},p'_1)\Gamma_k^S(p'_{2+3})
+\Gamma_k^S(p_{1+3})\lambda_k(p_2,p_{1+3};p'_{2+3},p'_1)\Gamma_k^S(p'_{2+3}) 
\nonumber\\
&&
+\Gamma_k^S(p_{2+3})\lambda_k(p_1,p_{2+3};p'_{1+3},p'_2)\Gamma_k^S(p'_{1+3})
-\Gamma_k^S(p_{1+3})\lambda_k(p_2,p_{1+3};p'_{1+3},p'_2)\Gamma_k^S(p'_{1+3}). \label{vac23}
\end{eqnarray}
Here we denote $p_{i+j}=p_i+p_j$. 
Indeed, we can show that this decomposition is possible directly from a formal diagrammatic expansion of six-point vertex in terms of the scattering length  $a_S$. 
When dimer exists, this vertex $\lambda_k$ represents scattering processes between atom and dimer. 

\begin{figure}[t]
\centering
	\includegraphics[width=35em]{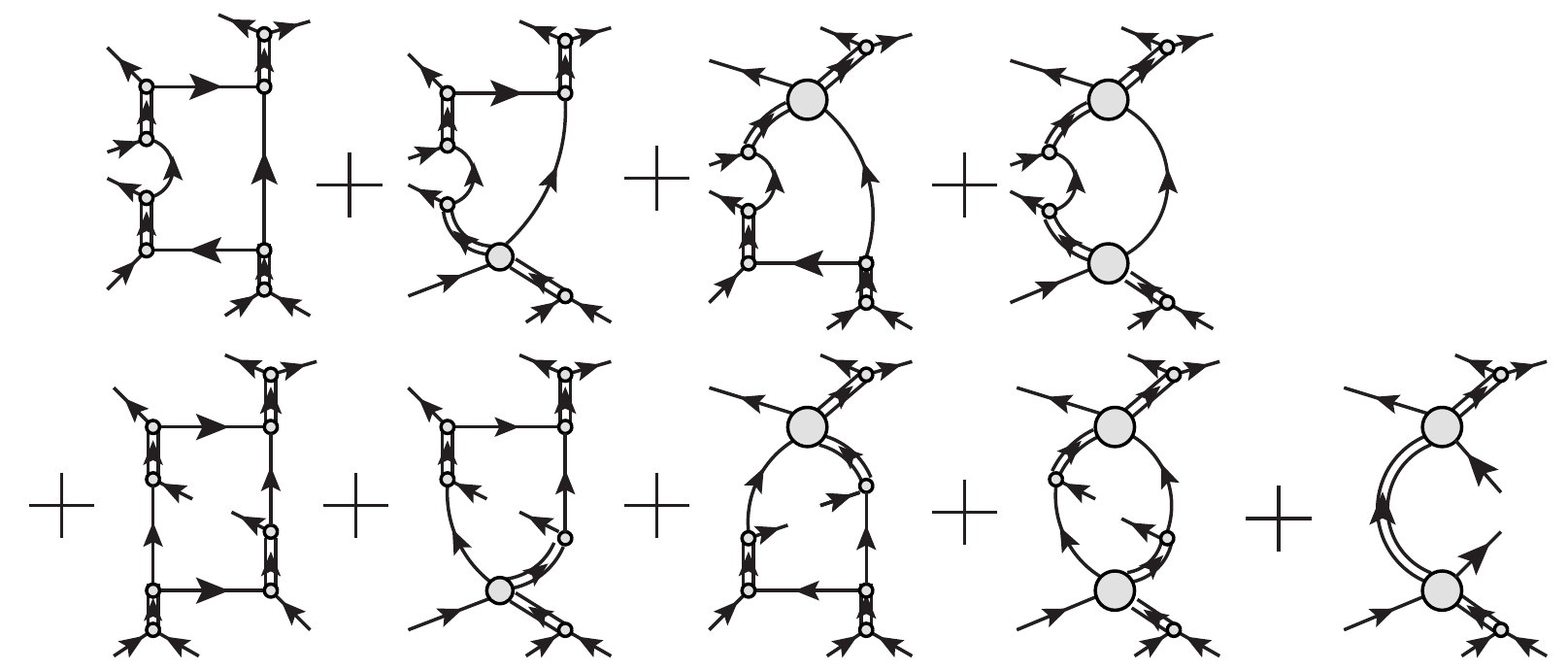}
\caption{All possible diagrams for the three-body sector in eight-point 1PI vertices. }
\label{ad8p}
\end{figure}

With this decomposition, we can extract the three-body sector in eight-point vertices, which can contribute the last diagram in the right hand side of Figure \ref{vac20}. 
All possible 1PI diagrams in the eight-point vertex, which can contribute as a feedback, are 
listed in Figure \ref{ad8p}. 
In order to obtain this result, we performed the perturbative expansion of the 1PI effective 
vertices in terms of the bare coupling $g$ and the fermion propagator $(G_0^{-1}+R_k)^{-1}$ 
at the scale $k$. After that, we took a resummation of the formal power series for 
the eight-point 1PI vertices in order to express them with 1PI vertices at the scale $k$ with fewer external legs. 
Notice that all these nine diagrams in Figure \ref{ad8p} are one-particle irreducible, with eight external lines, and belonging into three-body sector. 
The diagrams in Figure \ref{ad8p} are called ``1-closable'' diagrams in the paper \cite{Floerchinger:2013tp}. 

\begin{figure}[t]
\begin{equation*}
\parbox{11.5em}{\includegraphics[width=12em]{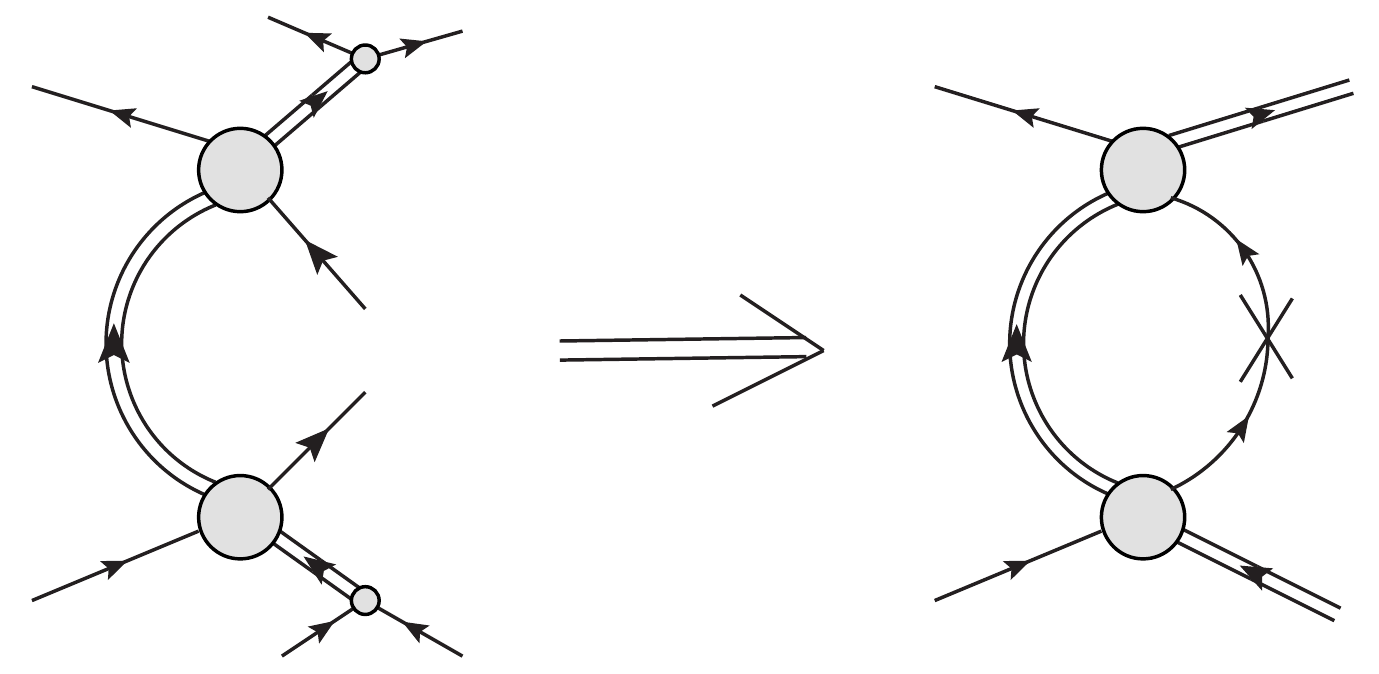}}\p_k R_k
\end{equation*}
\caption{Diagrammatic procedures to get feedback terms from diagrams in Figure \ref{ad8p}. 
\label{fig:amp}}
\end{figure}

The diagrammatic expression of eight-point vertices in Figure \ref{ad8p} gives an expression for the feedback term in the flow equation of $\lambda_k$. 
In order to obtain the feedback terms, we just have to do the following procedures (see also Figure \ref{fig:amp}): 
\begin{enumerate}
\item Amputate the external bosonic lines in each graph, and 
\item close two fermion lines with the two-point function $\p_k R_k$ so as not 
to make particle-hole loops. 
\end{enumerate}
This procedure is clearly possible in each diagram in Figure \ref{ad8p}, and then we here explicitly show that feedback from higher-point vertices is possible even in the vacuum. 
Furthermore, this procedure closes the hierarchy of the Wetterich equation (\ref{wet01}), 
and we need not consider four-body problems in discussing three-body scattering physics. 

\begin{figure}[t]
\begin{equation*}
\p_k \parbox{3em}{\includegraphics[width=3em]{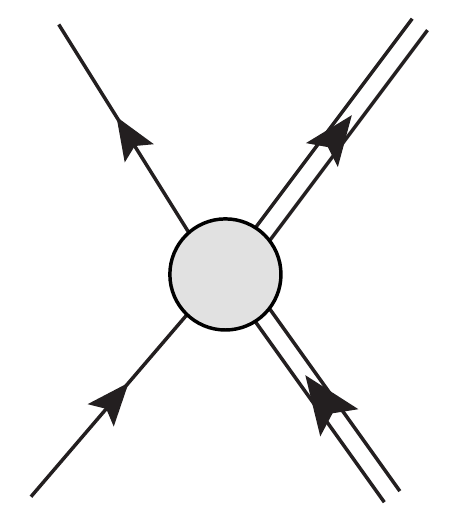}}
=\widetilde{\p}_k\left(
\parbox{15em}{\includegraphics[width=15em]{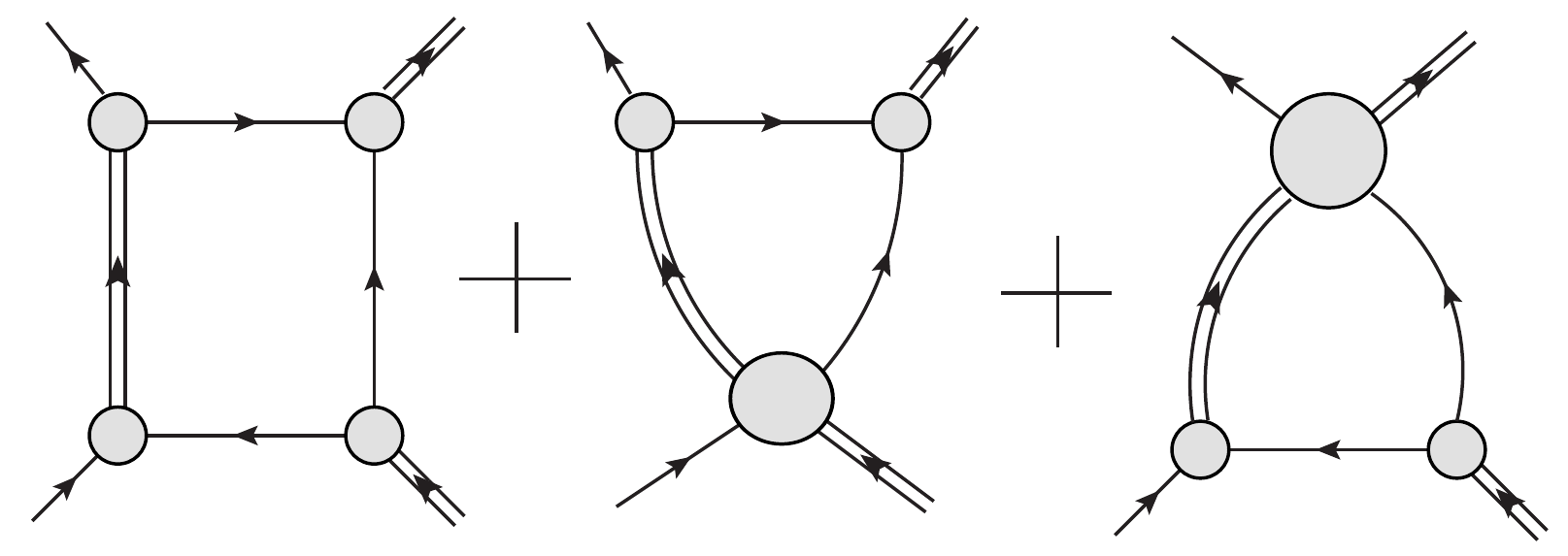}}
\right)
+\mbox{9 feedback terms}
\end{equation*}
\caption{Flow equation of $\lambda_k$. Here 9 feedback terms can be found from Figure \ref{ad8p}. }
\label{vac25}
\end{figure}

Diagrammatically, the flow equation for $\lambda_k$ can be represented as in 
Figure \ref{vac25}. 
``9 feedback terms'' in Figure \ref{vac25} are immediately obtainable from 
Figure \ref{ad8p} 
by following the procedure mentioned above. 
Using the momentum conservation, $P:=p_1+p_{2+3}=p'_1+p'_{2+3}$, we can reduce the number of variables in $\lambda_k$. 
Let us introduced a matrix notation so that matrix elements of $\lambda$, $G$, and 
$K$ are given by 
\begin{equation}
(\lambda)_{q,q'}=\lambda_k(P+q,-q;-q',P+q'),
\label{mat01}
\end{equation}
\begin{equation}
(G)_{q,l}=[G_0^{-1}+R_k]^{-1}(-l-P-q),
\label{mat02}
\end{equation}
and
\begin{equation}
(K)_{l',l}=(2\pi)^4\delta^4(l-l') {\Gamma_k^S(-l)\over [G_0^{-1}+R_k](l+P)}. 
\label{mat03}
\end{equation}
Here we omit the $k$-dependence just for simplicity of notation.
\begin{figure}[t]
\centering
$\lambda= \parbox{3em}{\includegraphics[width=3em]{fig7_1}}$, 
$G=  \parbox{5em}{\includegraphics[width=5em]{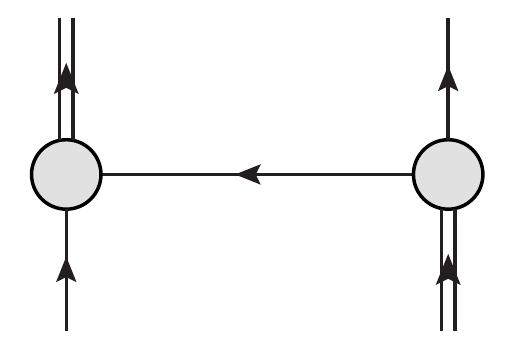}}$, 
$K= \parbox{5em}{\includegraphics[width=5em]{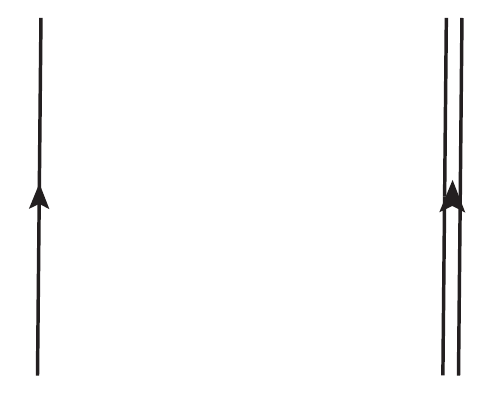}}$
\caption{Diagrammatic definition of symbols in (\ref{mat01}), (\ref{mat02}), and (\ref{mat03})}
\label{adf07}
\end{figure}

Using the matrix notation, we can write down the analytic expression of the flow equation 
in Figure \ref{vac25} as 
\begin{equation}
\p_k \lambda_k
=\widetilde{\p}_k (G\cdot K \cdot \lambda+\lambda\cdot K\cdot G+G\cdot K \cdot G)+
\mbox{9 feedback terms}. 
\label{ap01}
\end{equation}
From the diagrams in Figure \ref{ad8p}, 9 feedback terms are given as follows: 
\begin{eqnarray}
&&\mbox{9 feedback terms}\nonumber\\
&=&G\cdot (\p_k-\widetilde{\p}_k)K\cdot G+G\cdot K \cdot \p_k G\cdot K\cdot G
+G\cdot(\p_k-\widetilde{\p}_k)K\cdot \lambda 
\nonumber\\
&&+G\cdot K\cdot \p_k G\cdot K\cdot \lambda
+\lambda\cdot(\p_k-\widetilde{\p}_k)K\cdot G +\lambda\cdot K\cdot \p_k G\cdot K\cdot G
\nonumber\\
&&+\lambda\cdot(\p_k-\widetilde{\p}_k)K\cdot \lambda 
+\lambda\cdot K\cdot \p_k G\cdot K\cdot \lambda
+\lambda \cdot \widetilde{\p}_k K\cdot \lambda. \label{ap02}
\end{eqnarray}
Here we used the flow equation for the four-point coupling $\Gamma_k^S$, 
and then $(\p_k-\widetilde{\p}_k)K$ represent the self-energy correction of the dimer propagator 
in equation (\ref{ap02}). Substituting (\ref{ap02}) into (\ref{ap01}), we find that 
\begin{eqnarray}
\p_k \lambda_k&=&
\p_k (G\cdot K\cdot G)+\p_k(G\cdot K)\cdot \lambda+\lambda\cdot \p_k(K\cdot G)
\nonumber\\
&&+G\cdot K\cdot \p_k G\cdot K\cdot G+G\cdot K\cdot \p_k G\cdot K\cdot \lambda
\nonumber\\
&&+\lambda\cdot K\cdot \p_k G\cdot K\cdot G
+\lambda\cdot K\cdot \p_k G\cdot K\cdot \lambda
\nonumber\\
&&+\lambda \cdot {\p}_k K\cdot \lambda. \label{ap03}
\end{eqnarray}
This is the closed form of the flow equation for three-body scattering problems. 

In case of the two-body scattering problem with a contact interaction, 
such feedbacks are absent. 
This can be understood again most easily from diagrammatic considerations. 
The diagrams in six-point vertices do not have any structures like those in Figure \ref{ad8p}, 
because such six-point diagrams cannot be 1PI. Absence of the self-energy correction 
can also be understood from similar reasoning. 


\section{Formal solutions of three-body scattering problems\label{formal_sol}}

\begin{figure}[t]
\centering
$\parbox{3em}{\includegraphics[width=3em]{fig3_1}}
=\parbox{7em}{\includegraphics[width=7em]{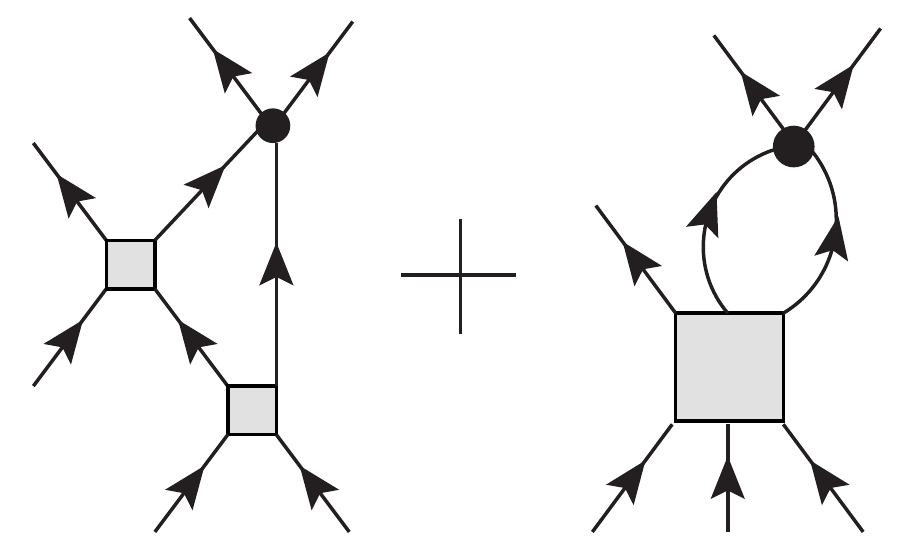}}$, \quad
$\parbox{3em}{\includegraphics[width=3em]{fig3_1}}
=\parbox{10em}{\includegraphics[width=10em]{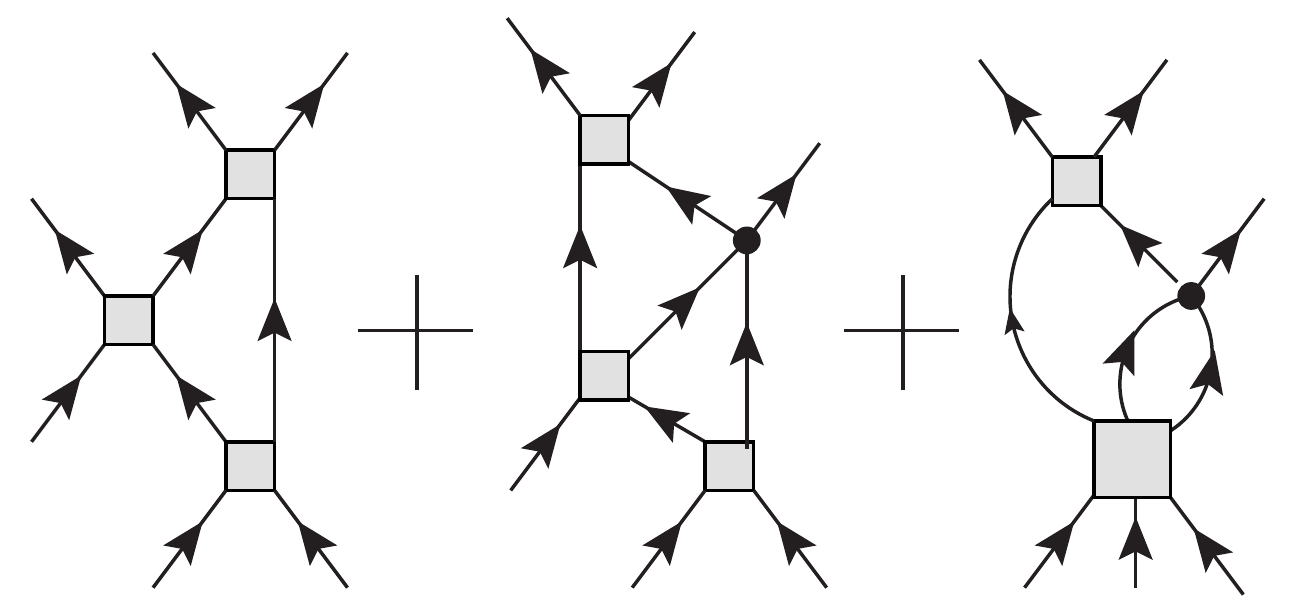}}$
\caption{Dyson-Schwinger equations for six-point 1PI vertices. 
Blobs represent bare couplings. }
\label{adf03}
\end{figure}

In order to find a formal solution for the atom-dimer scattering vertex $\lambda_k$, 
let us consider Dyson-Schwinger equations for the six-point vertex functions. 
We should point out that the Dyson-Schwinger equation must hold even for 
theories with an IR regulator $R_k$, and it can also be interpreted as an integrated flow 
of the Wetterich equation \cite{Pawlowski}. 
For the six-point vertex ${\Gamma_k}^{\up\up\down}_{\down\up\up}$ in our model, 
we have two Dyson-Schwinger equations for the scattering problem. 
Their diagrammatic expressions can be found in Figure \ref{adf03}. 
In deriving the second Dyson-Schwinger equation for the six-point vertex in 
Figure \ref{adf03}, 
we use the Dyson-Schwinger equation for the four-point vertex $\Gamma_k^S$. 
This second expression of the Dyson-Schwinger equation is equivalent 
to the integral equation for three-body scattering problems derived by S. Weinberg in 
\cite{PhysRev.133.B232}. 

\begin{figure}[t]
\begin{equation*}
\parbox{3em}{\includegraphics[width=3em]{fig7_1}}
=\parbox{8em}{\includegraphics[width=8em]{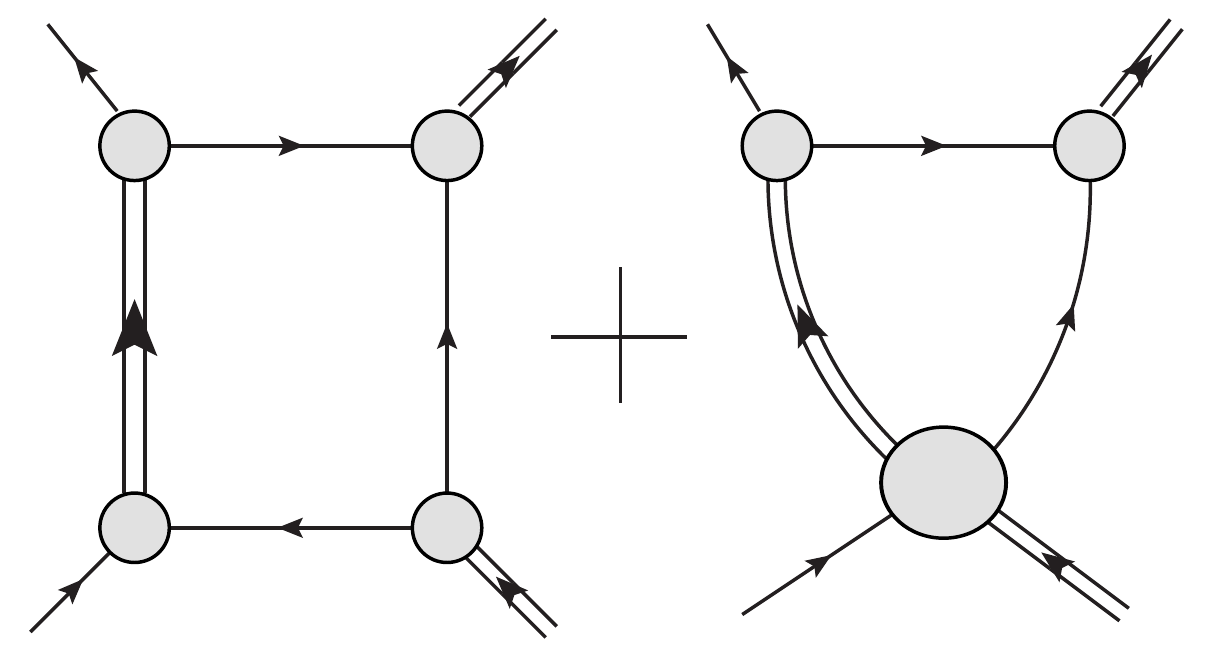}}
\end{equation*}
\caption{Diagrammatic expression of the integral equation for $\lambda_k$}
\label{adf05}
\end{figure}

Combining these two Dyson-Schwinger equations in Figure \ref{adf03} and applying 
the decomposition of the six-point 1PI vertex in (\ref{vac23}) or in Figure \ref{vac22}, 
we obtain the integral equation for the atom-dimer scattering vertex $\lambda_k$. 
Diagrammatically, $\lambda_k$ satisfies the integral equation in Figure \ref{adf05}, 
and its analytical expression is given by 
\begin{equation}
\lambda=G\cdot K\cdot \lambda+G\cdot K\cdot G. 
\label{adf01}
\end{equation}
At $k=0$, this integral equation is equivalent to the integral equation for the 
atom-dimer scattering $T$-matrix, which was originally derived by G. Skornyakov and 
K. Ter-Martirosyan \cite{skornyakov1957three}. 
We should emphasize that 
the integral equation (\ref{adf01}) holds at any values of the parameter $k$. 
Furthermore, the solution of this integral equation (\ref{adf01}) must satisfy the 
flow equation (\ref{ap03})

Let us explicitly show that the solution of this integral equation satisfies the flow equation 
(\ref{ap03}). 
We can formally write down the solution of (\ref{adf01}): 
\begin{equation}
\lambda
=G\cdot K \cdot G + G\cdot K \cdot G\cdot K \cdot G +\cdots. 
\label{adf08}
\end{equation}
To derive the flow equation, let us take a derivative with respect to $k$ 
of the both sides of (\ref{adf08}), which gives 
\begin{eqnarray}
&&\p_k \lambda\nonumber\\
&=&
\p_k (G\cdot K \cdot G) +\p_k(G\cdot K)\cdot (G\cdot K\cdot G+\cdots)
+(G\cdot K\cdot G+\cdots)\cdot \p_k(K\cdot G) \nonumber \\
&&\hspace{-1.em}
+(G+G\cdot K\cdot G+\cdots )\cdot K\cdot \p_k G \cdot K\cdot (G+G\cdot K\cdot G+\cdots)
\nonumber \\
&&\hspace{-1.em}
+(G\cdot K\cdot G+\cdots)
\cdot \p_k K\cdot 
(G\cdot K\cdot G+\cdots). \label{adf09}
\end{eqnarray}
Combining (\ref{adf08}) and (\ref{adf09}), we obtain 
\begin{eqnarray}
\p_k \lambda
&=&\p_k (G\cdot K \cdot G)+\p_k(G\cdot K)\cdot \lambda
+\lambda\cdot \p_k(K\cdot G)\nonumber\\
&&\hspace{-1.3em}+(G+\lambda)\cdot K\cdot \p_k G \cdot K\cdot (G+\lambda)
+\lambda\cdot \p_k K\cdot \lambda. 
\label{adf10}
\end{eqnarray}
Now the equation (\ref{ap03})
and the equation (\ref{adf10}) are the same, and the flow equation of FRG for few-body physics 
is explicitly shown to be equivalent to the Dyson-Schwinger equation. 
The procedure discussed here can provide a convenient way to obtain the closed flow equation 
of few-body physics. 

\section{Flow equation with auxiliary fields\label{auxiliary}}
So far, we have discussed the few-body physics using FRG only with fermions. In this section, 
we introduce an auxiliary field describing dimers and briefly discuss the flow of the atom-dimer 
scattering vertex. 
We consider the model specified by the classical Lagrangian  
\begin{equation}
\overline{\psi}G_f^{-1}\psi
+\phi^* G_b^{-1} \phi
+h\phi^*\psi_{\up}\psi_{\down}+h^*\phi\overline{\psi}_{\down}\overline{\psi}_{\up}, 
\label{aux01}
\end{equation}
where $G_f$ and $G_b$ are free propagators of $\psi$ and $\phi$, respectively, and 
$h$ is a coupling constant of the Yukawa-type interaction. 
Here $\psi$ is a Grassmannian field and $\phi$ represents a bosonic field. 
Lagrangian in (\ref{aux01}) is equivalent to the one in (\ref{cl_ac}) via the Hubbard-
Stratonovich transformation by putting $G_f^{-1}=(\p_{\tau}-\nabla^2/2m-\mu)$, 
$G_b^{-1}=-1/g$, and $h=i$. 
Since there are two different degrees of freedom $\psi$ and $\phi$ in this model, 
we introduce two regulating functions $R_k^{(f)}$ and $R_k^{(b)}$, so that 
the IR regulating term is given by 
\begin{equation}
\overline{\psi}R_k^{(f)}\psi+\phi^* R_k^{(b)} \phi. 
\label{aux02}
\end{equation}
These two regulating functions can be chosen independently, and we can go back 
to the model in the previous sections by putting $R_k^{(b)}=0$, which procedure 
corresponds to integration out of the bosonic degrees of freedom. 

\begin{figure}[t]
\centering
\includegraphics[width=0.35\textwidth]{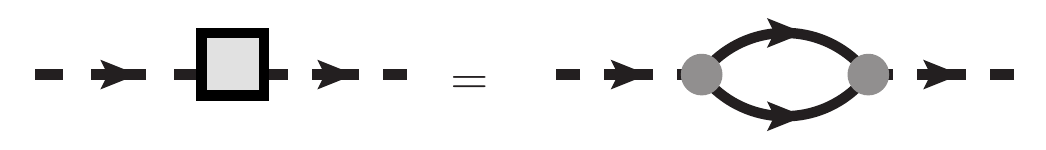}
\caption{Dyson-Schwinger equation for the dimer propagator. 
Here, dashed lines describe dimer propagators, the square box in the left hand side is 
the self-energy correction of dimer fields, the blobs in the write hand side are Yukawa couplings $h$, $h^*$, 
and the solid lines describe fermion propagators at the scale $k$: $(G_f^{-1}+R_k^{(f)})^{-1}$. 
As usual, external lines are amputated. }
\label{DS_dimer_prop}
\end{figure}

In order to derive the flow equation describing few-body physics of this model, 
we use the technology developed in section \ref{formal_sol}: we derive 
Dyson-Schwinger equations of the IR regularized theory at first and 
take the $k$-derivative of the formal solution of that integral equation. 
In this model, fermion propagators and the Yukawa coupling do not get quantum correction, 
and thus they do not flow under the change of $k$. 
Also, the 1PI vertex describing scattering between atoms remains zero at any $k$'s. 
On the other hand, bosonic dimer propagator acquires the quantum correction and its 
Dyson-Schwinger equation is given in Figure \ref{DS_dimer_prop}. 
This immediately gives the same flow equation given in Figure \ref{fig_fl4p}, 
and we find that the effective boson propagator is given by $-({\Gamma_k^S}^{-1}-R_k^{(b)})^{-1}$. 
Apart from the existence of $R_k^{(b)}$, we can identify the double lines of the diagrams in 
the previous sections with the dashed lines in this section. 

\begin{figure}[t]
\centering
\includegraphics[width=0.32\textwidth]{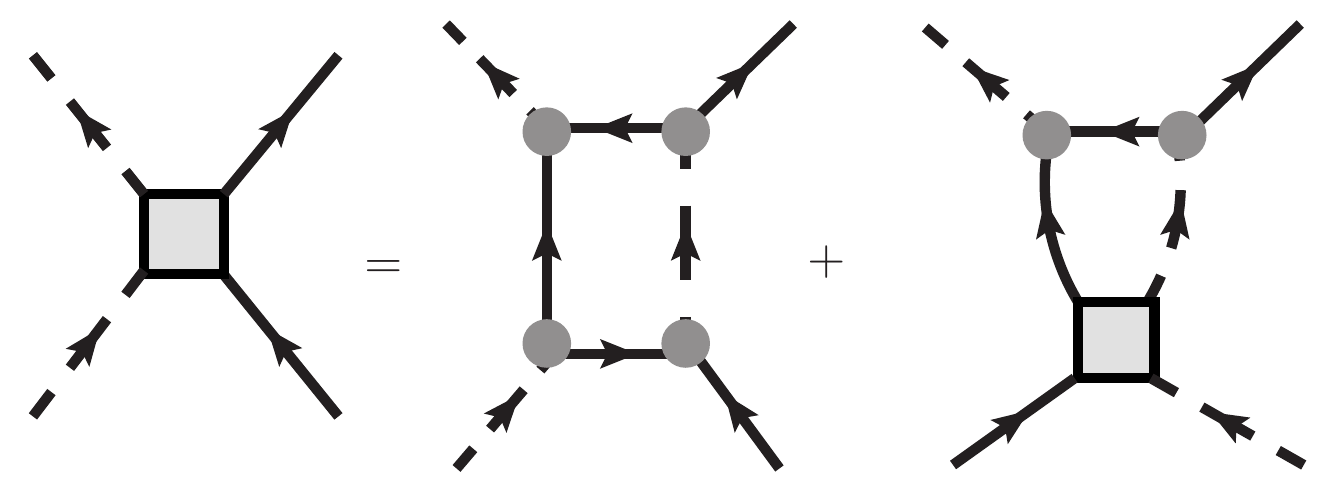}
\caption{Dyson-Schwinger equation for the atom-dimer 1PI vertex. }
\label{DS_atom_dimer}
\end{figure}

Now we can discuss scattering processes between an atom and a dimer using this model. 
The 1PI vertex involving one atom and one dimer satisfies the Dyson-Schwinger equation 
shown in Figure \ref{DS_atom_dimer}, which corresponds to the integral equation in Figure \ref{adf05}. 
Writing the atom-dimer 1PI vertex as $\lambda_k$ again, its formal solution is given by 
the series in (\ref{adf08}), in which $G$ and $K$ should be replaced by 
\begin{equation}
(G)_{q,l}=[G_f^{-1}+R_k^{(f)}]^{-1}(-l-P-q),
\label{aux03}
\end{equation}
and
\begin{equation}
(K)_{l',l}=(2\pi)^4\delta^4(l-l') {[{\Gamma_k^S}^{-1}-R_k^{(b)}]^{-1}(-l)\over [G_f^{-1}+R_k^{(f)}](l+P)}, 
\label{aux04}
\end{equation}
instead of (\ref{mat02}) and (\ref{mat03}). Now we find the flow equation of $\lambda_k$ for 
the model with auxiliary fields with any regulators, which is given by (\ref{adf10}). 

\begin{figure}[t]
\begin{eqnarray*}
-\p_k\lambda_k=\widetilde{\p}_k
\left(
\parbox{16em}{\includegraphics[width=0.4\textwidth]{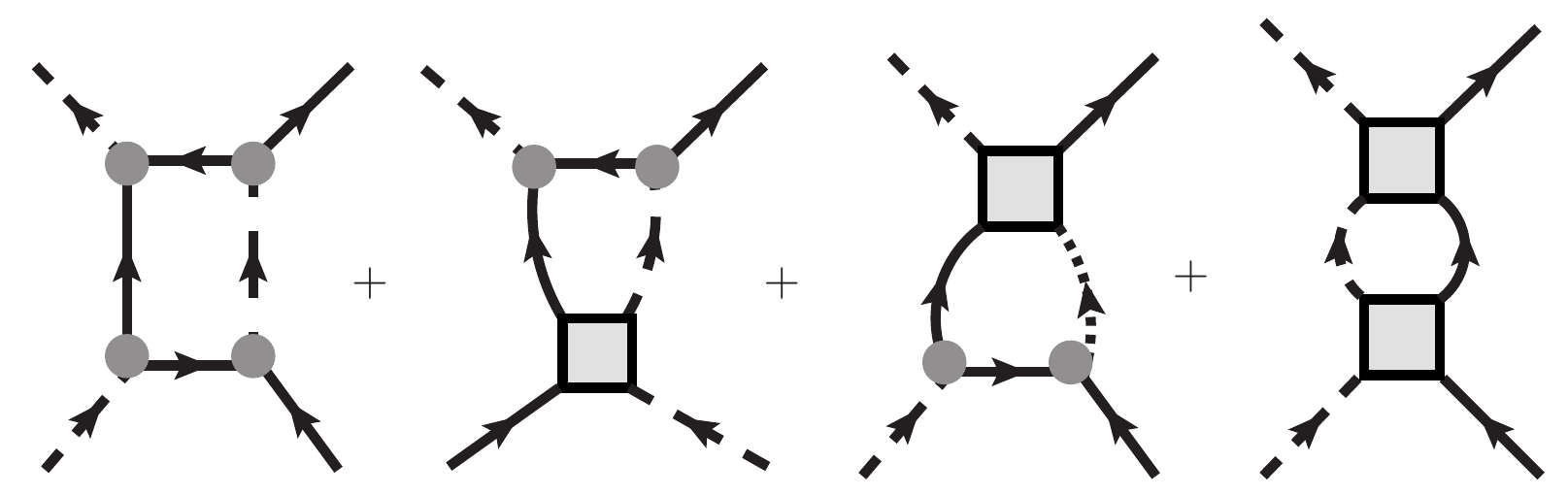}}
\right)+\parbox{4.em}{\includegraphics[width=0.1\textwidth]{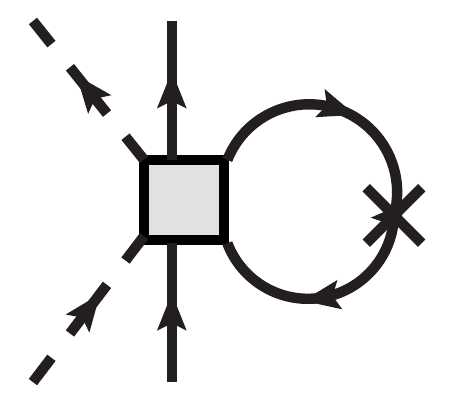}}{\p_k R_k^{(f)}}
\end{eqnarray*}
\caption{Flow equation of the atom-dimer 1PI vertex $\lambda_k$. }
\label{flow_ad_aux}
\end{figure}

Let us interpret this result from the viewpoint of Wetterich equation (\ref{wet01}). 
By separating the $k$-derivative $\p_k$ in the right hand side of (\ref{adf10}) into the 
derivative $\widetilde{\p}_k$ of $k$-dependence of regulators $R_k^{(f,b)}$ and the other parts, 
we find four terms, which can be directly obtained from the expansion of Wetterich equation, 
and other eight terms, which corresponds to feedback from higher-point vertices. 
Therefore, Wetterich equation of $\lambda_k$ takes the form given in Figure \ref{flow_ad_aux}, 
where the last feedback term completely corresponds to the first eight diagrams shown in 
Figure \ref{ad8p}. 
In the model with auxiliary fields, contribution from the last diagram in Figure \ref{ad8p} 
appears in the original one-loop contribution, but structure of other eight feedback terms 
is the same as before. 

\section{Conclusion}\label{conclusion}
In this paper, we showed that some one-loop diagrams in the flow equation, 
which look like particle-hole loops at first sight, 
do not vanish even for non-relativistic scattering problems in the vacuum. 
This can happen due to non-localities of the effective vertices, and 
this fact causes the hierarchy problem of the Wetterich equation for few-body physics. 
In order to construct the closed flow equation for those problems, 
we need to write down possible diagrams for higher-point vertices. 
We also suggest that Dyson-Schwinger equations can provide a convenient way 
to construct the closed flow equation, as discussed in Sections \ref{formal_sol} and 
\ref{auxiliary}. 

In order to construct a new flow equation of FRG for three-body physics, 
we identified all Feynman diagrams for eight-point vertices which contribute to the 
flow equation for three-body scattering problems. 
Combining the flow equation of FRG 
with diagrammatic considerations, we derived a closed flow equation 
for the three-body scattering problem. 
Furthermore, we proved consistency between our flow equation for the three-body problems 
and the corresponding Dyson-Schwinger equation, and the relation of our method 
with the Skornyakov and Ter-Martirosyan equation is unveiled. 
 
We also observed the structure of the flow equation when auxiliary dimer fields 
are introduced. Starting from the Dyson-Schwinger equation, we systematically 
derived the flow equation of FRG without any approximations and showed that 
even with the auxiliary fields there exists feedback from higher-point vertices for 
atom-dimer, or three-body, scattering problems. 

Let us compare our studies with previous works on few-body physics in FRG. 
In the paper \cite{PhysRevC.78.034001}, authors discussed the atom-dimer scattering 
with FRG by introducing bosonic fields via the Hubbard-Stratonovich transformation as in (\ref{aux01}). 
They showed that their flow equation for the atom-dimer scattering 
is consistent with the Skornyakov and Ter-Martirosyan equation if $R_k^{(f)}=0$ in (\ref{aux02}), 
in which fermions are integrated out at first so that only dynamical degrees of freedom 
are bosonic dimers. 
We here formally proved the consistency for more general cases, and, furthermore, 
we have constructed concrete procedures to treat three-body scattering problems using FRG. 

FRG has also been applied to few-body physics in order to reveal its universality, especially in the context of Efimov physics with the help of dimer fields \cite{PhysRevA.79.042705,floerchinger2011efimov,schmidt2012efimov}. 
In these studies, atoms are integrated out at first so that the Skornyakov and Ter-Martirosyan equation holds. In the context of renormalization group methods, Efimov physics can be interpreted as a limit cycle solution of the renormalization group transformations \cite{bedaque1999three}. 
Limit cycles can appear only when the flow equation of $\lambda_k$ depends on itself quadratically
\footnote{To see this fact clearly, let us treat $\lambda$ as a constant coupling. Then the flow equation of $\lambda$ takes the form $\p_k\lambda=A(k)\lambda^2+B(k)\lambda+C(k)$, where $A$, $B$, and $C$ are some coefficients. In order to realize a limit cycle, $\lambda$ must repeat to run from $-\infty$ to $+\infty$. Therefore, it is necessary that $B(k)^2-4A(k)C(k)<0$ in order for a limit cycle, but this is impossible if $A=0$.  }, 
and such quadratic terms of $\lambda_k$ appears through feedbacks from higher point vertices if dimer fields are not introduced. 
Therefore, in order to discuss Efimov physics using FRG without dimer fields, use of our formalism would be inevitable. 

We expect that a formulation for four-body scattering problems within FRG is 
also given in a similar way, but this task would be a future work. 
Even though we can solve such problems by deriving integral equations directly 
from Dyson-Schwinger equation, 
formulating that problem in the context of FRG is important 
for revealing many-body properties of physical systems, since FRG provides a unified 
description both for few-body and many-body physics.


\appendix
\section{Supplement for the flow equation of six-point vertex functions\label{app:6p}}
Analytic expression of the flow equation in Figure \ref{vac20} is given by 
\begin{eqnarray}
&&-\p_k {\Gamma_k}^{\up\up\down}_{\down\up\up}(p_1,p_2,p_3;p'_3,p'_2,p'_1)\label{vac21}\\
&&=
\widetilde{\p}_k\int_l\left(\scalebox{0.85}{$\displaystyle
{{\Gamma_k}^{\up\up\up\down}_{\down\up\up\up}(p_1,p_2,l,p_3;p'_3,l,p'_2,p'_1)
\over [G_0^{-1}+R_k](l)}
+{{\Gamma_k}^{\up\up\down\down}_{\down\down\up\up}(p_1,p_2,p_3,l;l,p'_3,p'_2,p'_1)
\over [G_0^{-1}+R_k](l)}$}\right)
\nonumber\\
&&
+\widetilde{\p}_k\int_l\left(\scalebox{0.85}{$\displaystyle
{\Gamma_k^S(p_{2+3}){\Gamma_k}^{\up\up\down}_{\down\up\up}(p_1,l,p_{2+3}-l;p'_3,p'_2,p'_1)
\over [G_0^{-1}+R_k](l)[G_0^{-1}+R_k](p_{2+3}-l)}
+{\Gamma_k^S(p_{1+3}){\Gamma_k}^{\up\up\down}_{\down\up\up}(l,p_2,p_{1+3}-l;p'_3,p'_2,p'_1)
\over [G_0^{-1}+R_k](l)[G_0^{-1}+R_k](p_{1+3}-l)}
$}\right. \nonumber\\
&&\left.+\scalebox{0.85}{$\displaystyle
{{\Gamma_k}^{\up\up\down}_{\down\up\up}(p_1,p_2,p_{3};p'_{2+3}-l,l,p'_1)\Gamma_k^S(p'_{2+3})
\over [G_0^{-1}+R_k](l)[G_0^{-1}+R_k](p'_{2+3}-l)}
+{{\Gamma_k}^{\up\up\down}_{\down\up\up}(l,p_2,p_3;p'_{1+3}-l,p'_2,l)\Gamma_k^S(p'_{1+3})
\over [G_0^{-1}+R_k](l)[G_0^{-1}+R_k](p_{1+3}-l)}$}\right) \nonumber\\
&&+\widetilde{\p}_k\int_l{1\over [G_o^{-1}+R_k](l)}\nonumber\\
&&\times\left(\scalebox{0.85}{$\displaystyle-
{\Gamma_k^S(p_{2+3})\Gamma_k^S(p'_{2+3})\Gamma_k^S(p_1+l)
\over [G_0^{-1}+R_k](l+p_{1-1'})[G_0^{-1}+R_k](p_{2+3}-l)}
+
{\Gamma_k^S(p_{1+3})\Gamma_k^S(p'_{2+3})\Gamma_k^S(p_2+l)
\over [G_0^{-1}+R_k](l+p_{2-1'})[G_0^{-1}+R_k](p_{1+3}-l)}$}\right.\nonumber\\
&&\left.\scalebox{0.85}{$\displaystyle+
{\Gamma_k^S(p_{2+3})\Gamma_k^S(p'_{1+3})\Gamma_k^S(p_1+l)
\over [G_0^{-1}+R_k](l+p_{1-2'})[G_0^{-1}+R_k](p_{2+3}-l)} 
-
{\Gamma_k^S(p_{1+3})\Gamma_k^S(p'_{1+3})\Gamma_k^S(p_2+l)
\over [G_0^{-1}+R_k](l+p_{2-2'})[G_0^{-1}+R_k](p_{1+3}-l)}$}
\right),\nonumber
\end{eqnarray}
where we have introduced the abbreviation $p_{2+3}=p_2+p_3$, $p_{1-1'}=p_1-p'_1$, etc. 
In section \ref{three_body}, we found physical meanings of each term when 
we consider atom-dimer scattering problems with the model given by (\ref{cl_ac}). 
The first two term in the right hand side of (\ref{vac21}) represents dissociation of dimers, 
and they become two-loop diagrams when we represent then using atom-dimer scattering 
vertex $\lambda_k$. Other terms in (\ref{vac21}) represent rescattering between 
atom and dimer. 

\section*{Acknowledgments}
The author thanks to Tetsuo Hatsuda for encouraging me to write this paper and 
for reading carefully manuscripts of this paper. 
The author is supported by JSPS Research Fellowships for Young Scientists.

\bibliographystyle{ptephy}
\bibliography{few_body_RG,FRG,BCS_BEC}

\begin{thebibliography}{10}

\bibitem{RevModPhys.80.1455}
Mark~G. Alford, Andreas Schmitt, Krishna Rajagopal, and Thomas Sch\"afer, Rev.
  Mod. Phys., {\bf 80}, 1455--1515 (Nov 2008).

\bibitem{PhysRevA.65.013606}
Peter Arnold, Guy Moore, and Boris Tom\'a\ifmmode~\check{s}\else \v{s}\fi{}ik,
  Phys. Rev. A, {\bf 65}, 013606 (Dec 2001).

\bibitem{PhysRevLett.83.1703}
Gordon Baym, Jean-Paul Blaizot, Markus Holzmann, Franck Lalo\"e, and Dominique
  Vautherin, Phys. Rev. Lett., {\bf 83}, 1703--1706 (Aug 1999).

\bibitem{bedaque1999three}
Paulo~F Bedaque, H-W Hammer, and U~Van~Kolck, Nuclear Physics A, {\bf 646}(4),
  444--466 (1999).

\bibitem{PhysRevC.77.047001}
Michael~C. Birse, Phys. Rev. C, {\bf 77}, 047001 (Apr 2008).

\bibitem{birse2005pairing}
Michael~C Birse, Boris Krippa, Judith~A McGovern, and Niels~R Walet, Physics
  Letters B, {\bf 605}(3), 287--294 (2005).

\bibitem{PhysRevA.83.023621}
Michael~C. Birse, Boris Krippa, and Niels~R. Walet, Phys. Rev. A, {\bf 83},
  023621 (Feb 2011).

\bibitem{Boettcher:2012cm}
Igor Boettcher, Jan~M. Pawlowski, and Sebastian Diehl, Nucl.Phys.Proc.Suppl.,
  {\bf 228}, 63--135 (2012).

\bibitem{diehl2010functional}
S.~Diehl, S.~Floerchinger, H.~Gies, JM~Pawlowkski, and C.~Wetterich, Annalen
  der Physik, {\bf 522}(9), 615--656 (2010).

\bibitem{PhysRevC.78.034001}
S.~Diehl, H.~C. Krahl, and M.~Scherer, Phys. Rev. C, {\bf 78}, 034001 (Sep
  2008).

\bibitem{PhysRev.186.456}
D.~M. Eagles, Phys. Rev., {\bf 186}, 456--463 (Oct 1969).

\bibitem{ellwanger1994flow}
U.~Ellwanger, Zeitschrift f{\"u}r Physik C Particles and Fields, {\bf 62}(3),
  503--510 (1994).

\bibitem{PhysRevA.79.013603}
S.~Floerchinger, R.~Schmidt, S.~Moroz, and C.~Wetterich, Phys. Rev. A, {\bf
  79}, 013603 (Jan 2009).

\bibitem{Floerchinger:2013tp}
Stefan Floerchinger, arXiv preprint arXiv:1301.6542 (2013).

\bibitem{floerchinger2011efimov}
Stefan Floerchinger, Sergej Moroz, and Richard Schmidt, Few-Body Systems, {\bf
  51}(2), 153--180 (2011).

\bibitem{PhysRevA.81.043628}
Boris Krippa, Niels~R. Walet, and Michael~C. Birse, Phys. Rev. A, {\bf 81},
  043628 (Apr 2010).

\bibitem{PhysRev.106.1135}
Tsin~D. Lee, Kerson Huang, and Chen~N. Yang, Phys. Rev., {\bf 106}, 1135--1145
  (Jun 1957).

\bibitem{leggett1980diatomic}
A.~Leggett,
\newblock Diatomic molecules and cooper pairs,
\newblock In {\em Modern Trends in the Theory of Condensed Matter}, volume 115
  of {\em Lecture Notes in Physics}, pages 13--27. Springer Berlin Heidelberg
  (1980).

\bibitem{PhysRevC.73.044309}
Masayuki Matsuo, Phys. Rev. C, {\bf 73}, 044309 (Apr 2006).

\bibitem{PhysRevA.79.042705}
S.~Moroz, S.~Floerchinger, R.~Schmidt, and C.~Wetterich, Phys. Rev. A, {\bf
  79}, 042705 (Apr 2009).

\bibitem{Morris1}
Tim~R. Morris, International Journal of Modern Physics A, {\bf 9}(14),
  2411--2450 (1994).

\bibitem{nozieres1985bose}
P.~Nozieres and S.~Schmitt-Rink, Journal of Low Temperature Physics, {\bf
  59}(3), 195--211 (1985).

\bibitem{Pawlowski}
Jan~M. Pawlowski, Annals of Physics, {\bf 322}(12), 2831--2915 (2007).

\bibitem{schmidt2012efimov}
R.~Schmidt, S.P. Rath, and W.~Zwerger, The European Physical Journal B, {\bf
  85}(11), 1--6 (2012).

\bibitem{skornyakov1957three}
GV~Skornyakov and KA~Ter-Martirosyan, Sov. Phys., JETP, {\bf 4}, 648--661
  (1957).

\bibitem{PhysRev.133.B232}
Steven Weinberg, Phys. Rev., {\bf 133}, B232--B256 (Jan 1964).

\bibitem{wetterich1993exact}
C.~Wetterich, Physics Letters B, {\bf 301}(1), 90--94 (1993).

\bibitem{Zwerger201110}
Wilhelm Zwerger, editor,
\newblock {\em The BCS-BEC Crossover and the Unitary Fermi Gas (Lecture Notes
  in Physics)},
\newblock  (Springer, 2012 edition, 10 2011).

\end{thebibliography}

\end{document}